\documentclass[preprint,aps]{revtex4}
\usepackage{graphicx}
\usepackage{mathptmx} % postscript fonts
\usepackage{natbib}

\begin{document}
%\hfill BNL-xxx       % for bnl number

\newpage

\title{
Point-occurrence self-similarity in crackling-noise systems
and in other complex systems
}
\author
{
\'Alvaro Corral  %%$^\dag$\email{alvaro.corral at uab.es},
}
\affiliation{
%%$^1$%
Centre de Recerca Matem\`atica,
Edifici Cc, 
Campus UAB,
E-08193 Bellaterra (Barcelona), Spain.\\
}
\date{\today}

\begin{abstract}
It has been recently found that a number of systems
displaying crackling noise also show a remarkable 
behavior regarding the temporal occurrence of successive events
versus their size:
a scaling law for the probability distributions of
waiting times as a function of a minimum size is fulfilled,
signaling the existence on those systems of self-similarity 
in time-size.
This property is also present in some non-crackling systems.
Here,
the uncommon character of the scaling law is illustrated
with simple marked renewal processes, built by definition with
no correlations.
Whereas processes with a finite mean waiting time 
do not fulfill a scaling law in general and tend towards a Poisson process
in the limit of very high sizes, 
processes without a finite mean tend to another class of
distributions, characterized by double power-law waiting-time
densities.
This is somehow reminiscent of the generalized central limit theorem.
A model with short-range correlations is not able to 
escape from the attraction of those limit distributions.
A discussion on open problems in the modeling of these
properties is provided.
\end{abstract}

%\pacs{
%???
%}% PACS, the Physics and Astronomy Classification Scheme.

\maketitle
\pagestyle{empty}

\section{Introduction}

In the words of Sethna {\it et al.}, 
``crackling noise arises when a system responds 
to changing external conditions through discrete, impulsive
events spanning a broad range of sizes'' \cite{Sethna_nature}.
Operationally, 
a broad range of sizes essentially means
that the size $s$ of the events fluctuates following a power-law distribution,
$D(s) \propto 1/s^{1+\beta}$, where $D(s)$ is the probability density of
$s$ and $1+\beta$ is the exponent.
This is quite remarkable, as power laws signal the absence of
characteristic scales, in this case of event sizes
\cite{Christensen_Moloney}.
It is also implicit that the %%action 
driving that makes the external conditions change 
is relatively small and smooth,
%%but enough to drive it out of equilibrium, 
quite different from the resulting bursty response;
therefore, 
crackling noise signals a highly nonlinear behavior.
%%this behavior is highly nonlinear.

Although these ideas have been developed within 
the physics of condensed matter \cite{Sethna_nature},
%%and many systems in this field display crackling noise, 
%% the %%geosciences 
natural hazards show perhaps the largest number and best illustrations
of crackling noise \cite{Malamud_hazards}, including
earthquakes \cite{Kanamori_rpp,Main_ng}, 
landslides and rock avalanches \cite{Malamud_hazards,Frette96},
%%tsunamis \cite{Burroughs},
volcanic eruptions \cite{Lahaie}, 
rainfall \cite{Peters_prl},
hurricanes \cite{Osso},
solar flares \cite{Arcangelis,Baiesi_flares},
the activity of the magnetosphere \cite{Wanliss},
and perhaps meteorite impacts \cite{Chapman}
(provided that the Earth moves ``slowly'' through space). %%% ???.
In other catastrophic phenomena, as 
forest fires \cite{Malamud_science,Corral_fires}
or the extinctions of biological species \cite{Raup},
the events span a broad range of sizes
but it is not clear if or in which cases they are power-law distributed.

Beyond the geosciences, notable examples of crackling noise
arise in physiology and human affairs, like
neuronal firings \cite{Beggs},
epileptic seizures \cite{Osorio}
and appearances of words in texts or speech \cite{Zipf1972}
(if the ``size of a word'' is measured by its rarity,
i.e., its position in a ranking of frequencies).
%% Faltan impresiones, emails, Struwitz: ojo!!
Note that most examples of crackling noise arise in systems
with a high degree of complexity, characterized by an enormous
number of degrees of freedom that interact between them.

%%In fact, 
Crackling noise can be considered as the most important and apparent 
property of systems displaying self-organized criticality (SOC).
%%For this kind of phenomenon, 
This concept goes one step beyond,
and proposes that
the origin of the power-law distribution of sizes 
(i.e., the hallmark of crackling noise)
is the existence of a critical point (analogous to those
of equilibrium phase transitions, which are well-known to have
scale-invariant properties)
to which the dynamics of the system is attracted
by means of a feedback mechanism that balances driving and dissipation
\cite{Bak_book,Jensen}.
The paradigmatic example is a sandpile over an open support
to which grains are slowly added:
when there are few grains, the pile is flat
and grain dissipation at the border is low, then the pile grows;
in contrast, when there are too many grains
they easily travel through the system and border dissipation is large,
so the slope of the pile decreases. 
At the end (in the attractor), the slope fluctuates 
around a critical value that balances the input and output of grains,
and this state should have scale-invariant properties, 
i.e., power-law statistics,
as in equilibrium critical phenomena.
The behavior of real-world sandpiles is more diverse than what
the SOC picture suggests, but that is a different story 
\cite{Feder_sand,Frette96}.
%%For our purposes, we are only interested in sandpiles as a metaphor.
Then, SOC is a plausible physical mechanism for the 
emergence of crackling noise,
but this does not preclude that other mechanisms
could also lead to crackling noise \cite{Sornette_critical_book}.

In practice, although it is very simple to test if a system
displays crackling noise (just measuring the size of the events and 
calculating their distribution, checking that the driving on the system
is slow and smooth),
it is not so easy to demonstrate the existence of SOC, 
as one would need to measure the relations between the internal
variables of the system and their fluctuations, 
and they should behave in the same way
as the equivalent ones in an equilibrium phase transition
\cite{Peters_np}.

\section{Scaling Law for Waiting-Time Statistics \label{sec2}}

%%In any case, %%note that 
Neither the above definition of crackling noise
nor the usual studies of SOC pay too much attention on how the 
discrete and impulsive response events that the system develops 
occur in time.
Is there a unique dynamical process that defines these behaviors?
And in that case, is it periodic? Is it chaotic? Is it random? 
How does the dynamics reflect the scale-invariant properties
of such systems? \cite{Bak_debates}.

The case of earthquakes exemplifies our poor understanding
of the dynamics of this kind of processes
\cite{Bak_debates,Mulargia_Geller}.
On the one hand, there is a widespread belief
in the notion of {\it characteristic earthquakes}:
the strongest events that a single fault segment is
able to generate are always almost the same
(same epicenter, same size, same focal mechanism)
and should occur at regular times
\cite{Stein_95,Stein_02}.
On the other hand, for extended regions, 
it is often assumed that mainshocks come 
in a total random way, i.e., from a Poisson process, 
and aftershocks 
follow a different process \cite{Gardner_Knopoff}.
Kagan has strongly argued against these simplistic views,
showing evidence of time clustering in earthquake occurrence,
just the behavior opposite to the characteristic-earthquake concept
\cite{Kagan_91,Kagan_Jackson_95}.
%%%???

Recently, it has been found that
some of the systems mentioned above as
prototypical of SOC or crackling noise display a remarkable
temporal behavior.
For such systems, let us consider the waiting time, also called
recurrence time, inter-event time or inter-occurrence time;
this is the time between consecutive events above a size threshold.
So, we take into account only events whose size $s$ verifies
$s \ge c$, where $c$ is the threshold value
(but notice that for instance in the case of earthquakes one does not distinguish
between foreshocks, mainshocks, and aftershocks). 
This defines a set of occurrence times, $t_i^c$,
denoting the occurrence of the $i$-th event above $c$,
from $i=0$ to $N_c$.
As each event is characterized 
by a unique occurrence time it is assumed that
the duration of the event is very short in the scale
of observation, and therefore the process can be described
as a stochastic {\it point process} \cite{Daley_Vere_Jones,Lowen}.
In addition, each event is also characterized
by its size, so the process can be considered
a {\it marked point process},
the size being a ``mark'' added to the time occurrence
(we do not consider spatial degrees of freedom
in this paper, but see Refs. \cite{Davidsen_distance,Corral_prl.2006}).
In any case, 
the waiting times for events with $s \ge c$ are obtained
straightforwardly as $\tau_i^c = t_i^c - t_{i-1}^c$,
with $i=1, 2 \dots N_c$.

The key element of analysis was introduced by Bak {\it et al.}
\cite{Bak.2002}, by means of a systematic study of
the statistics of $\tau_i^c$ as a function of the size threshold 
$c$. 
Although the rise of $c$ only eliminates some values of the
occurrence times, leaving the rest unaltered
(i.e., $t_i^c \rightarrow t_j^{c'}$, with $j \le i$ and $c < c'$), 
the waiting times are changed in a more complicated
way as they add in a variable number to give rise to the
larger (or not) new waiting times.
Usually, for the type of systems that have been studied so far,
the waiting times show a large variability, 
and the best characterization on these processes is by means
of the waiting-time probability density \cite{Corral_prl.2004}.

What has been found is that for many such systems,
these probability densities
verify a scaling law.
If for events with $ s\ge c$
we denote the waiting-time probability density by $D_c(\tau)$ 
and the mean waiting time by $\bar \tau_c$
($\bar \tau_c \equiv \int_0^\infty \tau D_c(\tau)d\tau$), 
the scaling law can be written as
\begin{equation}
D_c(\tau ) = F(\tau/\bar \tau_c) / \bar \tau_c,
\label{scaling}
\end{equation}
where $F$ is a scaling function, independent on $c$.
This means that the shape of the distribution is independent
on the scale given by $\bar \tau_c$ 
(which obviously is determined by the threshold $c$);
in other words, when the waiting time is measured
using as a unit its mean value, the results are
independent on the value of $c$, which 
implies the existence of a self-similarity in the process.
We will argue in the rest of this paper
that this is quite a remarkable result in general, 
difficult to justify with the use of simple stochastic models.

In the case crackling noise or SOC systems,
the mean waiting verifies 
%%$\bar \tau_c^{-1} =1/\lambda \propto 1/c^\beta$,
$\bar \tau_c \propto 
1/\int_c^\infty D(s) ds
\propto c^\beta$, if $\beta > 0$,
and substituting in the scaling law,
\begin{equation}
D_c(\tau) = \frac 1 {c^\beta} \tilde F\left(\frac \tau {c^\beta}\right),
\label{scaling2}
\end{equation}
where $\tilde F$ is the scaling function $F$ incorporating
the factor of proportionality between $\bar \tau _c$ and $c^\beta$.
Written in this form, the scaling law turns out to be a particular case of
the condition of scale invariance for functions with two variables,
$\tau$ and $c$ \cite{Christensen_Moloney}.
So, although for one variable the condition of scale invariance
yields a power law (for instance, for $s$ we have $D(s)\propto 1/s^{1+\beta}$),
for two variables, like $s$ and $\tau$, 
scale invariance leads to Eq. (\ref{scaling2}),
with $\tilde F$ an undetermined function.

Crackling-noise or SOC systems showing this behavior include 
earthquakes \cite{Corral_prl.2004,Corral_pre.2005}, 
fractures \cite{Astrom,Davidsen_fracture},
%%rainfall \cite{Peters_prl},
solar flares \cite{Baiesi_flares},
literary texts \cite{Corral_words},
or some paradigmatic sandpile models,
in contrast with previous belief \cite{Paczuski_btw,Laurson_upon}.
But this property is shared by other systems 
for which its crackling-noise nature is in doubt, as
printing requests in a computer network \cite{Harder_Paczuski},
forest fires \cite{Corral_fires} 
and 
tsunamis \cite{Geist_Parsons}
(although the latter seem to be power-law distributed \cite{Burroughs},
certainly they are not slowly driven;
rather, they are cracklingly driven by undersea 
earthquakes and landslides).
Even systems that do not crackle, 
as diverse climate records (temperature, river levels, etc.)
\cite{Bunde},
or systems for which the crackling behavior
is in the derivative of the response signal, 
as financial indices \cite{Yamasaki},
verify a scaling law as Eq. (\ref{scaling})
when the threshold is large enough that the events
above it become extreme events.
The corresponding scaling functions come in a variety of functional forms;
are there any preferred types?

\section{Models for  Time-Size Scale Invariance of Event Occurrence}

\subsection{Marked Poisson Process}

Which is the meaning and depth of the scaling law (\ref{scaling})?
Certainly, a {\it marked Poisson process} trivially fulfills it.
This is a marked point process in which the 
occurrence times follow a Poisson process, and
the sizes of the events (the ``marks'')
come from a random distribution independently 
on occurrence times and other sizes.
Its simulation is very simple, with independent identical
exponentially distributed waiting times and independent identical power-law distributed sizes
(in the case of crackling-noise systems).

Indeed, a Poisson process is completely characterized 
by its rate, let us say, 10 events per hour.
If we now raise the size threshold in such a way that
half of the events are eliminated and half of them 
survive, this is equivalent to a {\it random thinning}
of the events with a thinning probability equal to $1/2$
(in which any event has the same probability of 
being eliminated, independently of the rest),
due to the fact that the sizes of the events are 
uncorrelated.
So, we end with a Poisson process of rate equal to 
5 events per hour.

It is well known how to show this more rigorously.
Consider that events are removed
from a marked Poisson process with a probability $q$, and are kept 
with probability $p=1-q$; then, the probability that 
the number of events $N'$
that survive in a time interval of length $\Delta$ is equal to $k$ is,
$$
\Pr[N'=k] = \sum_{n=k}^\infty \Pr[N'=k | N=n] \Pr[N=n],
$$
where $\Pr$ denotes probability, 
``$|$'' conditional probability,
$N$ is the number of events in the interval prior to thinning,
and
$n$ counts all the possible values of $N$.
By hypothesis, the original process is Poisson of rate $\lambda$, so 
$Pr[N=n]= e^{-\nu} \nu^n / n!$, with $\nu \equiv \lambda\Delta$, 
and by the uncorrelated nature of the process
$\Pr[N'=k | N=n]$ is given by the binomial distribution, 
$$
\Pr[N'=k | N=n] = \left( \begin{array}{c} n \\ k\end{array}\right)
p^k q^{n-k}.
$$
Substituting both above,
$$
\Pr[N'=k ] 
= e^{-\nu} \frac{p^k}{k!}  \sum_{n=k}^\infty
\frac{q^{n-k}}{(n-k)!} \nu^n 
= \frac{(p\nu)^k}{k!} e^{-\nu} \sum_{n=k}^\infty
\frac{(q\nu)^{n-k}}{(n-k)!} 
= \frac{(p\nu)^k}{k!} e^{-\nu(1-q)}
= e^{-p\nu} \frac{(p\nu)^k}{k!},    
$$
which defines another Poisson process of rate $p\lambda$.
If we rescale the new rate $\lambda'=p\lambda$
as $\lambda' \rightarrow \lambda'/p$ we recover 
precisely the original Poisson distribution.
Note that $p$ is given by $p = \Pr[s \ge c' | s \ge c]$,
and in our context it turns out that
$p = (c/c')^\beta$.

So, could the trivial marked Poisson process explain
the scaling law Eq. (\ref{scaling})?
Certainly not, as none of the known examples
mentioned above are characterized by an exponential
scaling function.
We should go beyond this trivial explanation.

\subsection{Marked Renewal Process}

The shapes of the scaling functions found for the real data
mentioned above are rather diverse, 
including the gamma distribution, 
the stretched exponential, and the power law 
for large times.
It seems necessary to incorporate this shape 
into the point process modeling those systems.
The most straightforward way to do this is through 
a {\it renewal process}, which is characterized by independent identically
distributed waiting times, following a specific distribution
\cite{Daley_Vere_Jones}.
If we add to this model independent identically distributed sizes
we end with a process that we may call {\it marked renewal process}.
Note that there are no correlations whatsoever in this process,
but there is a memory of the last event
if the waiting-time distribution is not exponential.

The probability distribution of waiting times for events with 
$s \ge c'$ can be obtained from the probability distribution 
for those with $s \ge c$, if $c' \ge c$.
The idea is the same as in the previous subsection
but we will use the waiting-time representation
rather than the count-number representation of the process.
The same steps as in Ref. \cite{Corral_prl.2005}
will be followed,
although the case here is simpler.

We start using the survivor function, 
$S_{c'}(\tau)\equiv \Pr[\mbox{ waiting time $ > \tau $ for events with } s \ge c'] 
= \int_\tau^\infty  D_{c'}(\tau)d\tau$.
If an event of size $s_0 \ge c'$ has taken place,
the next one with $s \ge c'$ can happen in a variety of ways,
depending on the number of events with $c \le s < c'$ in between.
So, we can write,
$$
%%\begin{array}{l}
%%\Pr[\mbox{ waiting time $ > \tau $ for events } s \ge c']=
S_{c'}(\tau) =
\sum_{j=1}^\infty 
\Pr[\tau^{(j)} > \tau,\, s_1<c',\dots \, s_{j-1}<c',\, s_j\ge c']=\\
$$
$$
\sum_{j=1}^\infty 
\Pr[\tau^{(j)} > \tau \,| \,s_1<c',\dots \, s_{j-1}<c',\, s_j\ge c']
\, \cdot \Pr[s_1<c',\dots \, s_{j-1}<c',\, s_j\ge c'] =\\
$$
$$
\sum_{j=1}^\infty 
\Pr[\tau^{(j)} > \tau ]
\, \cdot \Pr[s_1<c']\cdots \Pr[s_{j-1}<c']\cdot \, \Pr[s_j\ge c'] =
%%\end{array}
%\label{P_eq}
$$
where $\Pr$ denotes probability, 
``$|$'' conditional probability, and the $j-$th return time
is defined, for events with $s\ge c$,
as $\tau^{(j)}_i=t^c_i-t^c_{i-j}$, that is, as the elapsed time between any event and
the $j-$th event after it 
(naturally, the first return time is the waiting time).
The conditions on $\Pr[\tau^{(j)} > \tau ]$ are eliminated
because waiting times are independent on sizes.
As in the previous subsection $ p \equiv \Pr[s \ge c' \,| \, s \ge c ] = \Pr[s \ge c' ]$
and $q\equiv 1-p=\Pr[s < c' \,| \, s \ge c ] = \Pr[s < c'] $
(the condition $s \ge c$ is always implicit, if it is not explicit).
Therefore,
$$
%%\begin{array}{l}
%%    \Pr[\mbox{ waiting time    $> \tau $ for events } s \ge c']=\\
S_{c'}(\tau) =
\sum_{j=1}^\infty 
p q^{j-1}\Pr[\tau^{(j)} > \tau ].
%%%\,| \,s_1<c',\dots s_{j-1}<c',s_j\ge c'] 
%%\end{array}
$$

If in this equation we derive with respect $\tau$ 
we obtain the probability densities of the return times; 
as the waiting times are considered independent on each other,
we use that the $j-$th-return-time distribution
is given by $j$ convolutions of the first-return-time distribution
(denoted by the symbol $*$) to get
\begin{equation}
    D_{c'}(\tau) = \sum_{j=1}^\infty 
p q^{j-1} [ D_c(\tau) ]^{*j} = 
   p D_c (\tau) +qpD_c (\tau)*D_c (\tau)
 +q^2p D_c (\tau)*D_c (\tau)*D_c (\tau)
   + \cdots
\label{eq5}
\end{equation}
where the exponent $*j$ means that $D_c (\tau)$
is convoluted with itself $j$ times. 
It is convenient to look at Eq. (\ref{eq5})
in Laplace space, where things are simpler, then
$D(s) \equiv \int_0^\infty e^{-s\tau} D(\tau) d \tau$,
which is a (moment) generating function, and
the convolutions turn into simple products,
%%%%,
%
\begin{equation}
   D_{c'}(s) =
   p D_c (s) 
\sum_{j=1}^\infty q^{j-1} [D_c (s)]^{j-1}
=     p D_c (s) 
   +q p D_c^2 (s)+
   q^2p D_c^3 (s)
   + \cdots
\end{equation}
% 
%Notice that
%we have used the same symbol $D$ for both the probability 
%densities and for their Laplace transforms
%(which we may call generating functions), 
%although they are different functions, of course.
As $q$ and $D_c(s)$ are smaller than one
(this is general for generating functions),
the infinite sum can be performed, turning out that
\begin{equation}
    D_{c'}(s) =
   \frac{p D_c (s)}{1- q D_c (s)}.
\label{transf1}
\end{equation}

Equation (\ref{transf1}) describes the 
effect of rising the threshold on the waiting-time distribution.
The next step is the scale transformation, which puts
the distributions corresponding to $c$ and $c'$ 
on the same scale. 
We will obtain this by removing the effect of the decreasing
of the rate, which, is proportional to $p$,
so,
\begin{equation}
   D_{c'}(\tau) \, \rightarrow \, p^{-1}  D_{c'}( \tau /p ),
\label{p1}
\end{equation}
and in Laplace space we get
\begin{equation}
   D_{c'}(s) \, \rightarrow \,  D_{c'}( p s).
\label{p2}
\end{equation}
Therefore, the combined effect of rising the threshold
plus rescaling leads to a transformation $\top$
that acts on the original distribution, 
\begin{equation}
   \top D_c(s) =
   \frac{p D_c (ps)}{1- q D_c(ps)}.
\label{transf2}
\end{equation}

We are very interested in 
the fixed points of this transformation, which are obtained by the solutions 
$D_c^*(s)$ of 
$$
\top D_c^*(s) = D_c^*(s),
$$
where $*$ now means fixed point;
The previous fixed-point equation 
is totally equivalent to the scaling law (\ref{scaling}).
Introducing the variable $\omega \equiv p s$ and substituting $p=\omega/s$
and $q=1-\omega/s$ in the fixed-point equation
we get, separating variables,
\begin{equation}
\frac 1 {sD_c^*(s)} - \frac 1 {s} =
\frac 1 {\omega D_c^*(\omega)} - \frac 1 {\omega} \equiv  
\frac 1 \lambda;
\label{separate}
\end{equation}
where we have made both functions equal to an arbitrary constant 
due to the fact that $p$ and $s$ are independent variables 
and so $s$ and $\omega$ are; 
then, the only way in which the equality could be fulfilled, 
for all $s$ and $\omega$, is that the function is a constant $1 /\lambda$.
The solution is then
\begin{equation}
D_c^*(s)=\frac 1 {1+ s/\lambda},
\label{PoissonLaplace}
\end{equation}
which is the Laplace transform of an exponential
distribution,
\begin{equation}
 D_c^*(\tau)=\lambda e^{-\lambda \tau}.
\end{equation}
The dependence on $c$ enters by means of $\lambda$,
as $\lambda^{-1} =\bar \tau_c $.
Note that this demonstration includes the one on the previous subsection, 
showing that the marked Poisson process displays a scaling law
for the waiting-time distribution, 
but in this case we have achieved a more general result, 
as the marked Poisson process is the only marked renewal process
which can fulfill such a scaling law
(when the rescaling is done with the mean waiting time $\bar \tau_c$).

We can go one step beyond and demonstrate that the marked Poisson
process is an attractor for the broad family of marked renewal processes
for which the mean waiting time $\bar \tau_c$ is finite.
The iterative application of transformation $\top$ with a finite
probability $p$ is equivalent to the limit $p \rightarrow 0$
in Eq. (\ref{transf2}).
Expanding that equation up to first order in $p$,
using $D_c(ps)  = 1 - \bar \tau_c p s + \cdots$,
yields
$$
\top D_c(s) = \frac{p D_c (ps)}{1- q D_c(ps)}  = 
\frac{p (1 -\bar \tau_c sp + \cdots )}{1- (1-p) (1 -\bar \tau_c sp + \cdots )}
 =
$$
$$
p \, \frac{ 1 -\bar \tau_c sp + \cdots }{1- (1-p  -\bar \tau_c sp + \cdots )}
 =
\frac{ 1 -\bar \tau_c sp + \cdots }{ 1  + \bar \tau_c s + \cdots }
 =
\frac{ 1 }{ 1  + \bar \tau_c s} + \cdots 
$$
which indeed corresponds to a Poisson process
when $p \rightarrow 0$.

This result illustrates the strange particularity of 
the scaling relation (\ref{scaling}):
among the infinite number of probability distributions 
with a finite mean that can define a marked renewal process,
only one type, the one with exponentially distributed waiting times,
fulfills the scaling law.
The results can be put in the language of the
renormalization group.
Indeed, the first part of the process, called thinning,
where the threshold is raised from $c$ to $c'$,
corresponds to a decimation of events. This is analogous
to the renormalization of the Ising model,
where some portion of the spins are eliminated
\cite{Kadanoff,Christensen_Moloney}.
The second part of the process correspond to a change of scale
in time, which is equivalent to the change of scale in 
real space renormalization. 
A third step, the renormalization of the field, is not 
necessary for the purposes of computing waiting-time statistics. 
So, the transformation $\top$ 
can be considered a renormalization transformation,
and we have seen how a renewal process (with a finite mean)
renormalizes into the trivial Poisson fixed point.
So, for all the renewal processes of this kind
(except for a set of zero measure, given by the Poisson process)
one expects a change under renormalization, and not
scale invariance.
This is one of the reasons why 
the existence of the scaling law (\ref{scaling})
is so intriguing.

\subsection{Marked Renewal Process without a Finite Mean Waiting Time}

What happens for renewal processes whose waiting-time density
does not have a finite mean? Obviously, a rate cannot be defined
as the inverse of the mean, nevertheless, still it is possible
to follow an approach that makes sense.
The first part of our transformation, in which the size threshold
is increased from $c$ to $c'$, does not change (\ref{transf1});
however, the rescaling with the mean cannot be applied.
We will use as a rescaling parameter $p = \Pr[s \ge c' \,| \, s \ge c]$,
but in contrast to the previous case we will raise $p$ 
to some power $r$ in Eqs. (\ref{p1}) and (\ref{p2}).
This is equivalent to seek for a scaling law
of the form
$$
    D_c(\tau)=R_c^r F(R_c^r \tau),
$$
where the rate $R_c$ is understood as the number 
of events per unit time in the time window under consideration.

The transformation $\top$ [Eq. (\ref{transf2})]
then becomes
\begin{equation}
   \top D_c(s) =
   \frac{p D_c (p^r s)}{1- q D_c(p^r s)}.
\label{transf3}
\end{equation}

In the same way as before, %% see Eq. (\ref{separate}), 
the fixed point equation leads to
\begin{equation}
\frac 1 {s^{1/r}D_c^*(s)} - \frac 1 {s^{1/r}} =
\frac 1 {\omega^{1/r} D_c^*(\omega)} - \frac 1 {\omega^{1/r}} \equiv  
a,
\label{separatetwo}
\end{equation}
whose solution is 
\begin{equation}
D_c^*(s) = \frac 1 {1 + a s^\alpha},
\label{fptwo}
\end{equation}
with $\alpha\equiv 1/r$.
Only for $0\le \alpha \le 1$, i.e., $r\ge 1$, this function 
represents a probability distribution;
this is so because 
for other values of $\alpha$ the expansion
of $D_c^*(s)$ does not correspond to the expansion
of a generating function.
%when the mean is finite, 
%the generating function goes, for small $s$ as
%$ 1 -\bar \tau s + \dots$
%and if not, it goes as
%$ 1 - a s^\alpha + \dots$.
However,  
when $0 < \alpha < 1$ a finite mean does not exist.

Indeed, for small $s$,
$D_c^*(s)=1 - a s^\alpha$;
this corresponds, if $0 < \alpha < 1$, to the Laplace transform 
of $D_c^*(\tau)=A/\tau^{1+\alpha}$, for large $\tau$, with 
$a \equiv -A \Gamma(-\alpha)$
and $\Gamma(-\alpha)$ the gamma function of $-\alpha$
\cite{Abramowitz}.
On the other hand, the behavior for large $s$ is
$D_c^*(s)= 1/(a s^\alpha)$, and by means of a Tauberian theorem
the limit behavior for small $\tau$ is 
$D_c^*(\tau)=1/(a \Gamma(\alpha)\tau^{1-\alpha})$.
Summarizing,
\begin{equation}
D_c^*(\tau) = \left\{ 
\begin{array}{ll}
\frac 1 {a \Gamma(\alpha) \tau^{1-\alpha}}, \, & \mbox{ for small } \tau, \\
\frac a {|\Gamma(-\alpha)|\tau^{1+\alpha}}, \, & \mbox{ for large } \tau. 
\end{array}
\right.
\label{double_power}
\end{equation}
So, two power laws coexists, with exponents  $1-\alpha$ and $1+\alpha$.

Which is the basin of attraction of the fixed-point distribution,
$D_c^*(\tau)$?
Let us consider $D_c(\tau) \simeq A / \tau^{1+\alpha}$
for large $\tau$,
with $ 0 < \alpha < 1$.
Among the class of functions that do not have a finite mean, 
these are by far
the most important.
We already know that
the Laplace transform of $D_c(\tau)$ behaves as
$D_c(s) = 1 + A \Gamma(-\alpha) s^{\alpha} + \dots$,
for $s \rightarrow 0$ 
and so,
$D_c(p^r s) = 1 -a p^{r\alpha} s^{\alpha} + \dots$.
%%and  $a \equiv - A \Gamma(-\alpha)$.
Substituting into the new equation for $\top D_c(s)$,
and taking into account that $p\rightarrow 0$,
$$
\top D_c(s) = \frac{p D_c (p^r s)}{1- q D_c(p^r s)}  = 
\frac{p (1 -a s^\alpha p^{r\alpha} + \cdots )}{1- (1-p) 
(1 -a s^\alpha p^{r\alpha} + \cdots )}
 =
$$
$$
p \, \frac{ 1 - a s^\alpha p^{r\alpha}  + \cdots }
{1- (1-p  - a s^\alpha p^{r\alpha} + \cdots )}
 =
\frac{ 1 -a s^\alpha p^{r\alpha}  + \cdots }{ 1  + a s^\alpha p^{r\alpha-1}   + \cdots }
 =
\frac{ 1 }{ 1  + a s^\alpha} + \cdots 
$$
where we have used that $r \equiv 1/\alpha$, 
which, remember, means that the rescaling depends on the power-law
exponent of the waiting-time density.
Rescaling in this way ensures the existence of 
an attractor for the waiting-time densities that behave as
a power law for long times,
and this attractor is given by the fixed point (\ref{double_power}).
%We could also have verified that the previous expression
%is a solution of the fixed point
%equation corresponding to the new transformation (\ref{transf3}),
%proceeding in the same way to arrive to an equation 
%that is a generalization of Eq. (\ref{separate}),
%but the previous result is more powerful.

Why does this counterintuitive rescaling provide the fulfillment of a
scaling law? The reason is in the generalized central limit theorem.
Imagine that $p=1/2$, so we remove half of the events;
in order to recover the original pattern (characterized by the same
waiting-time probability density) we need to multiply the time
interval under consideration not by $1/p =2$ but by $1/p^{1/\alpha}$, which is larger than
$1/p$ for $ 0 < \alpha < 1$.
This is due to the fact that as the process evolves in time, 
longer waiting times appear, due to their power-law tail,
being necessary to consider far longer time intervals.

Reference \cite{Bouchaud_Georges} puts numbers into this 
simple intuitive explanation.
The total time interval up to the $N-$th event is
$t_N = \sum_{n=1}^N \tau_n$. 
The largest waiting time $\tau_m(N)$ among the $N$ values 
of the waiting time
can be estimated as $N \int_{\tau_m(N)}^\infty D_c(\tau) d\tau \simeq 1$,
this yields $\tau_c(N) \propto N^{1/\alpha}$ for large $N$,
and this means that during the time interval of lenght $t_N$
the process ``does not see'' the tail of $D_c(\tau)$ beyond
$\tau_m(N)$ and one can effectively truncate the distribution at
this value. 
Then, the ``typical'' value of $t_N$ can be associated
to the mean value of the truncated distribution, 
so $t_N \simeq N \langle \tau \rangle_{trunc} \propto N^{1/\alpha}$,
and in this way the total time up to the $N-$th event scales
in the same way as the largest waiting time.

Indeed, the generalization of the central limit theorem 
introduces rigor in this argument,
see for instance Ref. \cite{Bouchaud_Georges}.
If the $\tau_n$'s are power-law distributed, 
with a tail $D_c(\tau) = A/\tau^{1+\alpha}$, the standard central limit theorem
does not hold for $t_n$ and one has to apply its generalization.
Rescaling $t_N$ as $Z_N \equiv t_N/ N^{1/\alpha}$
the theorem states that the variable $Z_N$ 
has as a limit distribution, for $N\rightarrow \infty$,
one of the so called Levy stable distributions
(whose mathematical form is not relevant for our purposes).

Coming back to our problem,
there is a particular value of $\alpha$
for which the exact inverse Laplace transform of the 
fixed-point distribution can be easily obtained.
Indeed, if $\alpha=1/2$, we get that 
%%for $p \rightarrow 0$
$$
D_c^*(s) = \frac{ 1 }{ 1  + a \sqrt{s}};
$$
whose inverse Laplace transform turns out to be
\begin{equation}
D_c^*(\tau) =
\frac 1 {a \sqrt{\pi \tau}} 
- \frac{e^{\tau/a^2}}{a^2}\mbox{erfc}\left(\frac{\sqrt{\tau}}{a}\right),
\label{solution}
\end{equation}
where erfc(x) is the complementary error function,
erfc$(x) = 2 \pi^{-1/2} \int_x^\infty e^{-x^2} dx$
\cite{Abramowitz}.
Its asymptotic expansion will be useful
to learn how $ D_c^*(\tau)$ behaves,
$$
\mbox{erfc}(x) = \frac {e^{-x^2} }{\sqrt{\pi} \, x} 
\left(1- \frac 1{2 x^2} + \dots \right),
$$
this leads to 
$$
D_c^*(\tau) \simeq \frac a {2 \sqrt{\pi} \, \tau^{3/2}} = \frac A  {\tau^{3/2}},
$$
for large $\tau$,
with $a=-A\Gamma(-1/2)= 2 \sqrt{\pi} A$.
This is indeed coincident with the tail of the original distribution.
On the other hand, for small argument, erfc$(x) \simeq 1$
and then
$$
D_c^*(\tau) \simeq \frac 1 {2 A \pi \sqrt{\tau}},
$$
for short times, in agreement with the previous result for
a general value of $\alpha$, Eq. (\ref{double_power}).

\begin{figure*}
\centering
{\Large \hspace{-8cm} a}\\
\includegraphics[height=7.5cm]{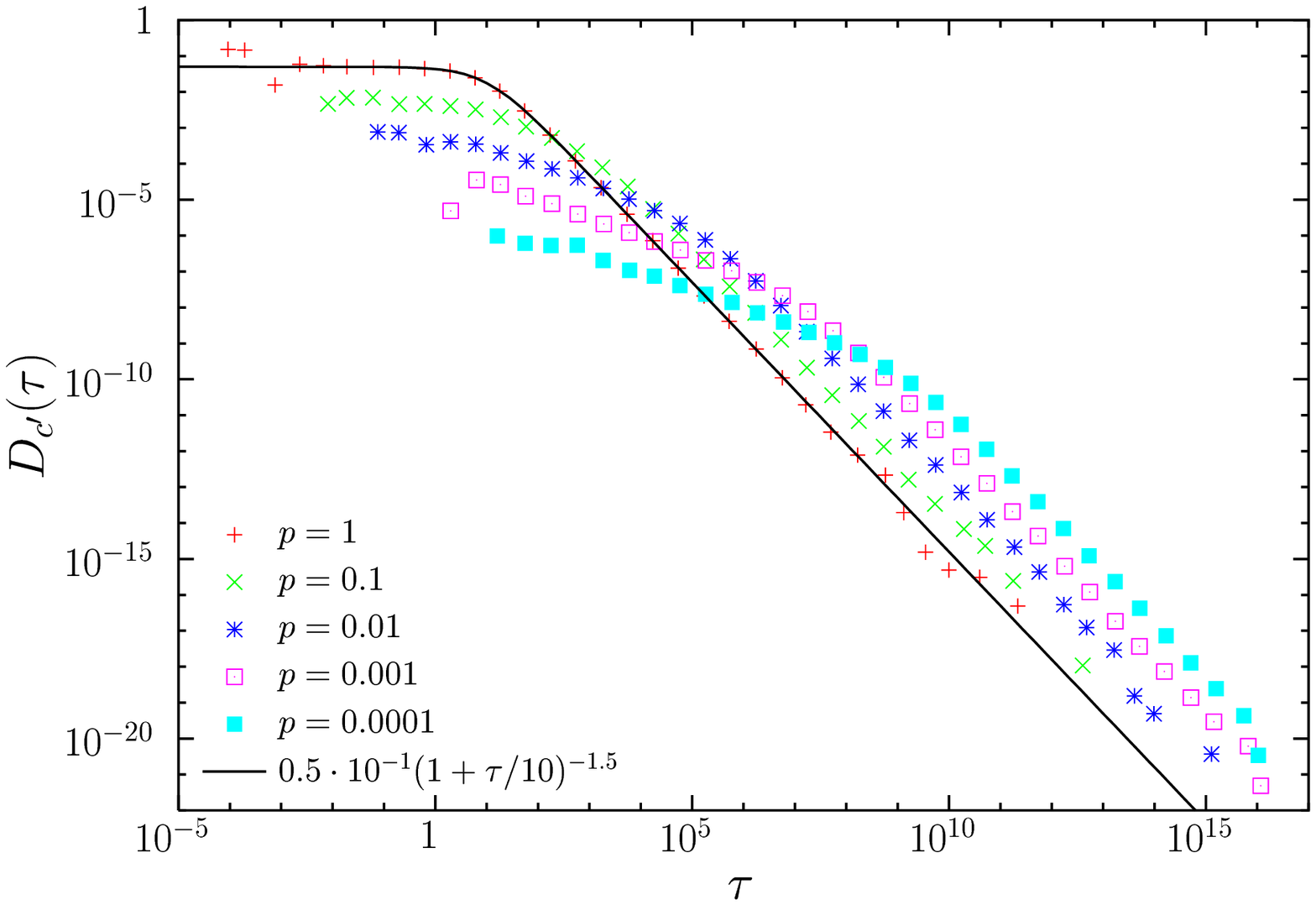}\\
{\Large \hspace{-8cm} b}\\
\includegraphics[height=7.5cm]{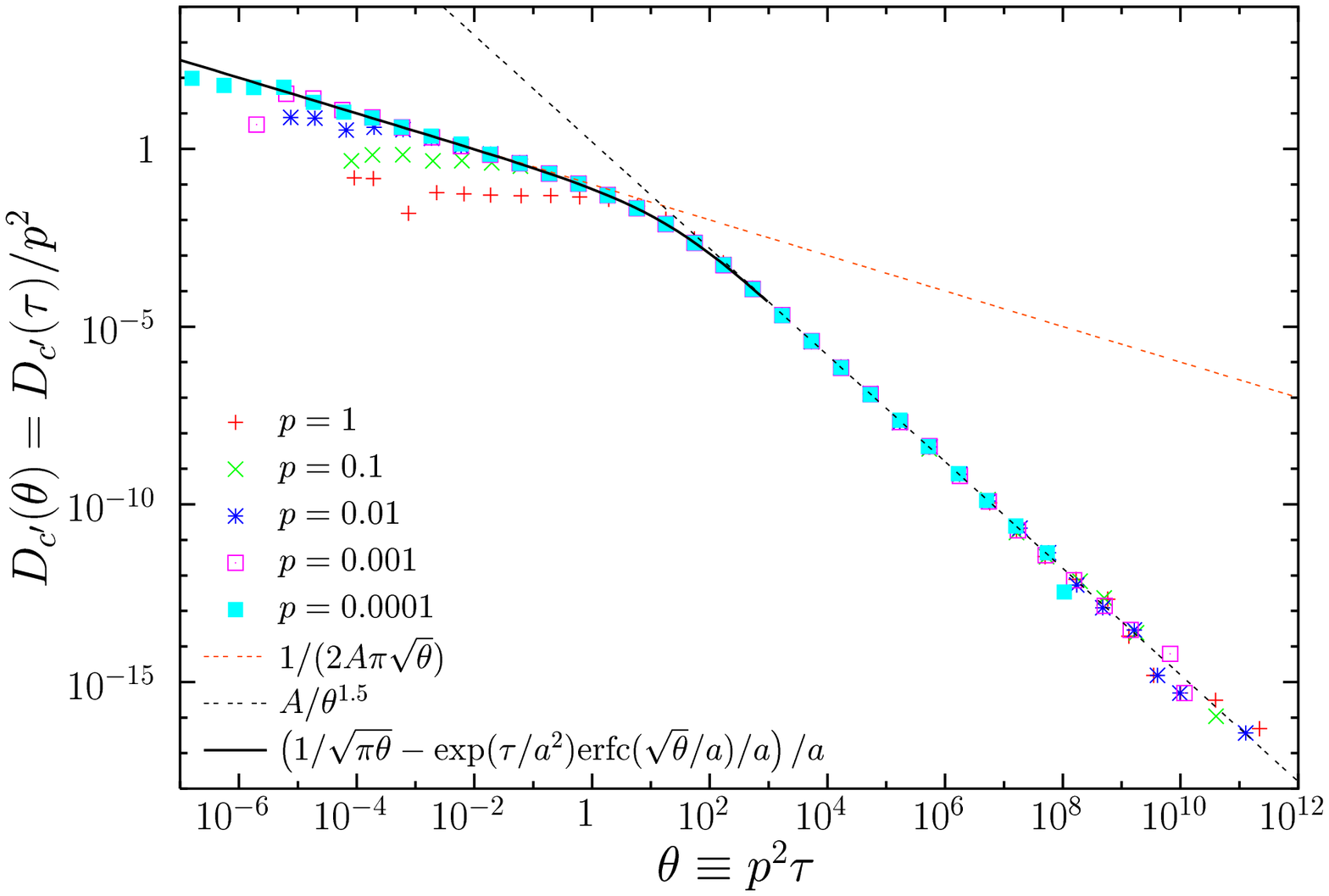}\\
\caption{
Illustration of the thinning plus rescaling transformation.
A marked renewal process is simulated, with 
waiting-time distribution given by Eq. (\ref{original}),
using $\alpha=0.5 $ and $\ell=10$.
(a) The effect of thinning, which removes events with probability $1-p$
is shown for different values of $p$.
(b) The complete transformation $\top$, adding rescaling by $p^2$ to thinning,
shows how the resulting distribution approaches the fixed-point solution,
Eq. (\ref{solution}).
}
\label{Dtau} 
\end{figure*}

We can test these results simulating the process
for instance for
\begin{equation}
D_c(\tau) = \frac \alpha  
{\ell \left( 1 + \tau/\ell\right)^{1+\alpha}}
\label{original} 
\end{equation}
for which $A=\alpha \ell^\alpha$.
The results appear in Fig. \ref{Dtau}.
Figure \ref{Dtau} (a)
shows this distribution, and the 
distributions that result after 
rising the size threshold,
keeping 10 \%, 1 \%, etc. up to 0.01 \% of the events.
Figure \ref{Dtau} (b) displays the same distributions
but including the rescaling, 
which allows to see the complete effect of
the transformation (\ref{transf3})
for different values of $p$.
It is clear how for $p=0.01$ the distribution 
is very close to the expected fixed-point
distribution (\ref{solution}),
and the agreement improves for smaller $p$.

Can these results be useful for crackling-noise systems or other complex systems?
Solar flares, when they are close to the minimum of the solar
cycle \cite{Baiesi_flares}, and also 
e-mail activity from individuals \cite{Goh}
and human-movement episodes \cite{Struzik}
%% no!!! %%% printing requests in a computer network
show power-law distributed waiting times,
with exponents $1+\alpha$ about 1.5, 1, and 1.8, respectively.
However, there are no indications of a second power law
with exponent $1-\alpha$, as requested by our theoretical calculations.

Nevertheless, these are peculiar systems;
first, in the case of solar flares, the thresholds
that define the waiting-time distributions are not
size thresholds but intensity thresholds
(size is defined as the integral of intensity over time),
and it is not clear how this change can modify the properties of 
the system.
In addition, the scaling which defines the scaling law is done
with the mean rate, as it would correspond to a distribution 
with a well-defined mean.
Second, for human movements and e-mail activity, the scaling law is not achieved
by means of a thinning of the process through the increasing of a
threshold, 
rather, individuals with different rates are compared.
And as a fourth example we could consider the BTW sandpile model, 
whose behavior (and the approach with which it has been studied)
is very similar to that of solar flares,
with a waiting-time exponent around 1.7 \cite{Paczuski_btw}.

\subsection{Processes with Short-Range Correlations}

Reference \cite{Corral_prl.2005}
introduced a very simple point process in which
each waiting time was correlated with the 
size of the previous event, in such a way that 
waiting times following larger events were drawn
from a waiting-time density with a short characteristic
time and waiting times after small events had 
a longer characteristic time.
The transformation of the waiting time density
when the threshold is increased and the time
is rescaled
verifies an equation which is a generalization
of the previous cases,
$$
\top D_c(s) = \frac{p D_{c\uparrow}(p^r s)}
{1-D_c(p^r s) + p D_{c\uparrow}(p^r s)}
$$
where $D_{c\uparrow}( s)$ is the Laplace transform
of $D_{c\uparrow}(\tau)$, which is the probability 
density of the waiting time that follows an event of size $c' > c$; 
more precisely $D_{c\uparrow}(\tau) \equiv D_c(\tau_i | s_{i-1} \ge c')$.

As in the last subsections, for $p \rightarrow 0$ we
can write
$D_c(s) = 1 - a s^\alpha +\dots$
and
$D_{c\uparrow}(s) = 1 - a' s^{\alpha '} +\dots$,
where $a'$, which gives the scale of the distribution,
may depend on $p$, increasing as $p \rightarrow 0$.
Substituting in the equation for the transformation $\top$,
$$
\top D_c(s)  = 
\frac{p (1 -a' s^{\alpha'} p^{r\alpha'} + \cdots )}{1- (1 - a s^\alpha p^{r\alpha} +\dots)
+p (1 -a' s^{\alpha'} p^{r\alpha} + \cdots )}
 =
$$
$$
p \, \frac{ 1 - a' s^{\alpha'} p^{r\alpha'}  + \cdots }
{a s^\alpha p^{r\alpha} +  \cdots +p (1 -a' s^{\alpha'} p^{r\alpha} + \cdots )}
 =
\frac{ 1 -a' s^{\alpha'} p^{r\alpha'}  + \cdots }{ 1  + a s^\alpha p^{r\alpha-1}   + \cdots }
 =
\frac{ 1 }{ 1  + a s^\alpha} + \cdots, 
$$
with, as usual, $r\equiv 1/\alpha$;
so, whatever the dependence of $a'$ on $p$, the shape
of $D_{c\uparrow}(\tau)$ is totally irrelevant to determine
the asymptotic behavior of the process, 
provided that $a'p \rightarrow 0$ when $p\rightarrow 0$.
This demonstrates that for the simple model we are considering, 
short-range correlations are not enough to escape from the 
attraction of the renewal fixed-point distributions.
For the case of processes with a finite mean, a demonstration
was already provided by Molchan, but the author believes 
the one here is more direct \cite{Molchan_private}.

\section{  %%Renormalization-Group Transformation and
Discussion}

We have seen how, when a mean waiting time exists, a process without correlations
cannot account for the appearance of a scaling law for waiting-time
distributions 
(except in the trivial case of a marked Poisson process, 
which is not observed in real systems). 
In the same way,
when a process without correlations
is characterized by the absence of a finite
mean waiting-time,
the theoretical results do not compare well
with observations either.
Therefore, correlations build the shape 
of the waiting-time distribution in crackling-noise
as well as non-crackling noise systems
(at least for the cases studied so far, 
see Sec. \ref{sec2}).

But short-range correlations %%as introduced in Ref. \cite{Corral_prl.2005},
do not seem  enough to break the dominance of the trivial Poisson 
fixed point when a mean exists, or the double power-law distributions
expected when the mean does not exist, 
as we have shown for the simple example introduced in Ref. \cite{Corral_prl.2005}.
As in equilibrium critical phenomena \cite{Binney}, long-range correlations
should be necessary  in order to escape from the basin of attraction
of the trivial fixed point.
Further research is necessary regarding this issue,
both from a fundamental point of view 
and with the goal of finding 
stochastic models of the systems displaying scaling laws.

A promising approach is that of Lennartz {\it et al.} \cite{Lennartz},
where a long-range correlated series of magnitudes is generated,
associating each magnitude value to a discrete time
(the authors have in mind earthquakes, but the results are more general).
Then, only extreme events, i.e., events above a large magnitude threshold, 
are considered, and the corresponding waiting-time statistics is obtained,
with a surprising agreement with observational earthquake data \cite{Corral_prl.2004}.
In other words, starting with a delta distribution of waiting times
for the initial process, renormalization leads to a nontrivial fixed point.
It is an open question why this is so.

The fact that a nontrival (nonexponential) scaling law for waiting-time distributions
may exist has been criticized by Molchan \cite{Molchan} and Saichev and Sornette
\cite{Saichev_Sornette_times}.
Assuming that seismic occurrence is well described by the ETAS model, 
the latter authors were able to derive the form of the 
waiting-time probability density.
Here we just mention that the ETAS (epidemic type aftershock sequence) model 
is the simplest modeling of the process of earthquake triggering that
puts all earthquakes on the same footing: any earthquake triggers 
other events with a probability that is proportional to two main factors:
the Omori law, which controls the decay in time of the seismic rate,
and the productivity law, which links the rate with the magnitude of the
triggering earthquake.
In addition, the magnitudes of the resulting triggered earthquakes 
are drawn independently from
the Gutenberg-Richter distribution.
Despite its simplicity, the mathematical treatment of the ETAS model
becomes an authentic {\it tour de force}.
In any case, Saichev and Sornette get that $D_c(\tau)$
fulfills something like a scaling relation, Eq. (\ref{scaling}),
but with a scaling function which is not a only function
of the rescaled waiting time,  
$$
F(x,\epsilon) = \left[ \frac{n \theta \epsilon^\theta }{x^{1+\theta}}+
\left ( 1-n + \frac{n\epsilon^\theta}{x^\theta}\right)^2
 \right]
\exp \left (- (1-n) x - \frac n {1-\theta} \epsilon^\theta x^{1-\theta} \right).
$$
where
$1+\theta$ is the exponent of the decay of the rate with time, 
given by the Omori law ($\theta \simeq 0$),
$n < 1$ is the branching parameter,
defining the number of events triggered directly by any event,
%%%$h \equiv n/(1-\theta)$,
and
$\epsilon = \lambda C$,
with
$C$ the small time constant that avoids divergence at zero time
in the Omori law.
So, in addition to the dependence on the rescaled waiting time
$x=\lambda\tau$, the density depends on the rate $\lambda$ through $\epsilon$,
in contrast with the idea of scale invariance.

The fact that the ETAS model does not fulfill an exact scaling
law does not seem highly surprising, after all,
due to the fact that the ETAS model itself is not fully 
self-similar \cite{ Vere_Jones,Saichev_Sornette_Vere_Jones}. 
At the end, for vanishing rate, $\epsilon \rightarrow 0$,
it turns out that $F(x)$ tends to an exponential distribution, 
and then the ETAS model renormalizes into the trivial Poisson process.
It would be of the maximum interest to calculate if 
real self-similar models, as the Vere-Jones model
\cite{Vere_Jones,Saichev_Sornette_Vere_Jones}
or the Lippiello-Godano-de Arcangelis model
\cite{Lippiello}
fulfill the scaling relation $(\ref{scaling})$
and which is the corresponding scaling function.
This is of course an unsolved problem.

As a final comment, let us mention that the models used in this
paper are purely stochastic, or mathematical.
Some readers, however, could ask for a more physical approach.
For the author,
the situation is analog to the study of diffusion processes
using random-walk models \cite{Bouchaud_Georges}:
the outcome is robust and independent on small details about the nature
of interactions 
(molecular collisions in one case and event-event triggering mechanisms
in the other).
Nevertheless, there are examples of crackling systems 
that have been successfully modeled on physical grounds, 
using for instance the random field Ising model \cite{Sethna_nature},
the so called ABBM model \cite{abbm,Zapperi_np},
or models of dislocation dynamics in plastic deformation
\cite{Csikor}.
It would be of the maximum interest to explore 
the time structure of events in those models.
Regarding the geosciences, which are the systems
we have in mind for this paper, the situation is more complicated, as 
the physics of those phenomena is still poor understood, and
controlled experiments cannot be performed.
It is a great challenge to find microscopic models
of natural catastrophes that give rise to the self-organized
structures that emerge in the long-run limit in those systems.

%It is noteworthy that Saichev and Sornette have provided more refined versions
%of their approach, which include the effect of long-range earthquake
%triggering between different regions.

The author thanks M. Bogu\~n\'a for his quick guide to Abelian
and Tauberian theorems.

%%\bibliographystyle{plain}  % alfabetico
%%%%\bibliographystyle{apalike}
%\bibliographystyle{unsrt}   % por orden de cita
%%\bibliographystyle{nature}

%\bibliography{../biblio}

\newpage

\end{document}